\documentclass[aps,pra,twocolumn,superscriptaddress,showkeys]{revtex4-1}

\usepackage{graphicx}
\usepackage{amsthm,amssymb,amsmath}
\usepackage{textcomp}
\usepackage{color}
\usepackage{braket}
\usepackage{gensymb}
\usepackage[euler]{textgreek}
\usepackage{float}
\usepackage{multirow}

\begin{document}

\title{Configurable heralded two-photon Fock-states on a chip}

\author{Xin Hua}
\address{Universit\'e C\^ote d'Azur, CNRS, Institut de Physique de Nice (INPHYNI), UMR 7010, Parc Valrose, 06108 Nice Cedex 2, France.}
\author{Tommaso Lunghi}
\address{Universit\'e C\^ote d'Azur, CNRS, Institut de Physique de Nice (INPHYNI), UMR 7010, Parc Valrose, 06108 Nice Cedex 2, France.}

\author{Florent Doutre}
\address{Universit\'e C\^ote d'Azur, CNRS, Institut de Physique de Nice (INPHYNI), UMR 7010, Parc Valrose, 06108 Nice Cedex 2, France.}
\author{Panagiotis Vergyris}
\address{Universit\'e C\^ote d'Azur, CNRS, Institut de Physique de Nice (INPHYNI), UMR 7010, Parc Valrose, 06108 Nice Cedex 2, France.}
\author{Gr\'egory Sauder}
\address{Universit\'e C\^ote d'Azur, CNRS, Institut de Physique de Nice (INPHYNI), UMR 7010, Parc Valrose, 06108 Nice Cedex 2, France.}
\author{Pierrick Charlier}
\address{Universit\'e C\^ote d'Azur, CNRS, Institut de Physique de Nice (INPHYNI), UMR 7010, Parc Valrose, 06108 Nice Cedex 2, France.}
\author{Laurent Labont\'e}
\address{Universit\'e C\^ote d'Azur, CNRS, Institut de Physique de Nice (INPHYNI), UMR 7010, Parc Valrose, 06108 Nice Cedex 2, France.}
\author{Virginia D'Auria}
\address{Universit\'e C\^ote d'Azur, CNRS, Institut de Physique de Nice (INPHYNI), UMR 7010, Parc Valrose, 06108 Nice Cedex 2, France.}
\author{Anthony Martin}
\address{Universit\'e C\^ote d'Azur, CNRS, Institut de Physique de Nice (INPHYNI), UMR 7010, Parc Valrose, 06108 Nice Cedex 2, France.}
\author{Sorin Tascu}
\address{Research Center on Advanced Materials and Technologies, University Alexandru Ioan Cuza, Romania.}
\author{Marc P. De Micheli}
\address{Universit\'e C\^ote d'Azur, CNRS, Institut de Physique de Nice (INPHYNI), UMR 7010, Parc Valrose, 06108 Nice Cedex 2, France.}
\author{S\'ebastien Tanzilli}%
\address{Universit\'e C\^ote d'Azur, CNRS, Institut de Physique de Nice (INPHYNI), UMR 7010, Parc Valrose, 06108 Nice Cedex 2, France.}
\author{Olivier Alibart}
\email{Olivier.Alibart@univ-cotedazur.fr}
\address{Universit\'e C\^ote d'Azur, CNRS, Institut de Physique de Nice (INPHYNI), UMR 7010, Parc Valrose, 06108 Nice Cedex 2, France.}

\date{\today}

\begin{abstract}
Progress in integrated photonics enables combining several elementary functions on single substrates for realizing advanced functionnalized chips. We report a monolithic integrated quantum photonic realization on lithium niobate, where nonlinear optics and electro-optics properties have been harnessed simultaneously for generating heralded configurable, two-photon states. Taking advantage of a picosecond pump laser and telecom components, we demonstrate the production of various path-coded heralded two-photon states, showing 94\% raw visibility for Hong-Ou-Mandel interference. The versatility and performance of such a highly integrated photonic entanglement source enable exploring more complex quantum information processing protocols finding application in communication, metrology and processing tasks.
\end{abstract}


\maketitle

\section{Introduction}
Entanglement stands as a key resource in quantum information science, being widely exploited to outperform classical communication, metrology, and computation protocols~\cite{Thew:2019}. Over a decade of theoretical and experimental efforts, the field of optical quantum
information processing~\cite{OBrien:07} has led to processors that may solve problems that classical computers cannot~\cite{Lund:17}, while embryonic quantum networks can distribute entanglement over continental distances~\cite{Pan:18}. Photonic qubits are seen as key candidates due to weak decoherence, the possibility of coding entanglement over several observables, and the availability of efficient and robust optical components for routing and manipulating the generated photons.
In this framework, integrated photonics stands as an enabling tool for realizing complex and scalable quantum circuits~\cite{Metcalf:13} otherwise unfeasible using bulk approaches. Solutions exploiting photonic devices relying on arrays of many photon sources have been shown to be useful for speeding-up quantum sampling algorithms~\cite{brod:19}, or for efficiently approximating on-demand sources of single photons~\cite{Bonneau:15,Collins:13}. The integration of several identical high-performance photon sources~\cite{Varnava:15} as well as the entangling circuitry for general purpose quantum processing is one the key challenges. From the practical side, strategic efforts are devoted for developing photonic chip showing low energy cost, high-efficiency internal photon sources, and fast and convenient phase modulation for enabling multiple high-fidelity quantum operations on single chips. 

Several platforms have achieved notable success for building efficient and versatile sources of photonic entanglement~\cite{Guo:17,Francesconi:20,Kim:20,Atzeni:18, Vergyris:16} associated with complex circuits with many linear components on the same substrate~\cite{meany:14,Jin:14,Martin:12,O'Brien:09}. Among the available material platforms, silicon photonics has recently drawn attention for its superior integration density factor~\cite{Faruque:18,Paesani:20} despite its lack of natural electro-optical coefficient leading to slow optical modulation via local heating. On the other hand, lithium niobate (LN) waveguides are known for their superior optical performance, such as low optical transmission losses, large second-order nonlinear coefficients (optics-optics as well as electro-optics) making it a serious contender for the fabrication of top-notch integrated photonics devices for both classical and quantum information applications~\cite{Alibart:16}. On the opposite to spontaneous four-wave mixing in silicon, quantum states can be generated in LN with unparalleled brightness by spontaneous parametric down-conversion (SPDC), further tailored and modulated by domain engineering simplifying the pump rejection~\cite{Zhu:12,Kruse:15}. Electro-optic modulation~\cite{Luo:19} bandwidth has reached up to 100\,GHz on LN on insulator chip\cite{Wang:18} opening the route to high-speed single photon switching rates. It is therefore natural to expect, with new LN technologies, to realize advanced quantum circuits fulfilling the requirements of next-generation quantum photonics integration.

Here, we demonstrate the generation of configurable heralded path-coded two-photon states in the telecom C-band out of a single LN chip showing unprecedented level of integration. Our chip merges all the necessary functions for the generation and the deterministic separation of two non-degenerate photon-pairs in a heralded single photon configuration as well as Hong-Ou-Mandel (HOM) interference on a tunable optical coupler. The performances are characterized by a 94\% raw visibility and a heralded entanglement flux of $\sim 200$\,Hz. Switching from a two-photon separate state $|1,1\rangle$ to a N00N state $\frac{1}{\sqrt{2}}\left(|2,0\rangle+|0,2\rangle\right)$ is also demonstrated through electro-optic phase shift (EOPS). Our results highlight the potential of monolithic integration of photon-pair sources and interferometric functions for practical quantum technology applications. Such low energy cost, flexible photon sources, fast and convenient phase modulators, and reconfigurable waveguide circuits find application in discrete and continuous variable quantum optics~\cite{Lenzini:18,Mondain:19} but also in hybrid quantum optics~\cite{Morin:14}.

We first present the operating principle of the chip, then we report the classical characterizations for each part, and eventually provide the results obtained with this device in the quantum regime and demonstrate its capacity to generate high-quality heralded N00N state.

\section{Operating principle of chip and classical characterization}

\begin{figure*}[!htp]
\centering
\includegraphics[width=\linewidth]{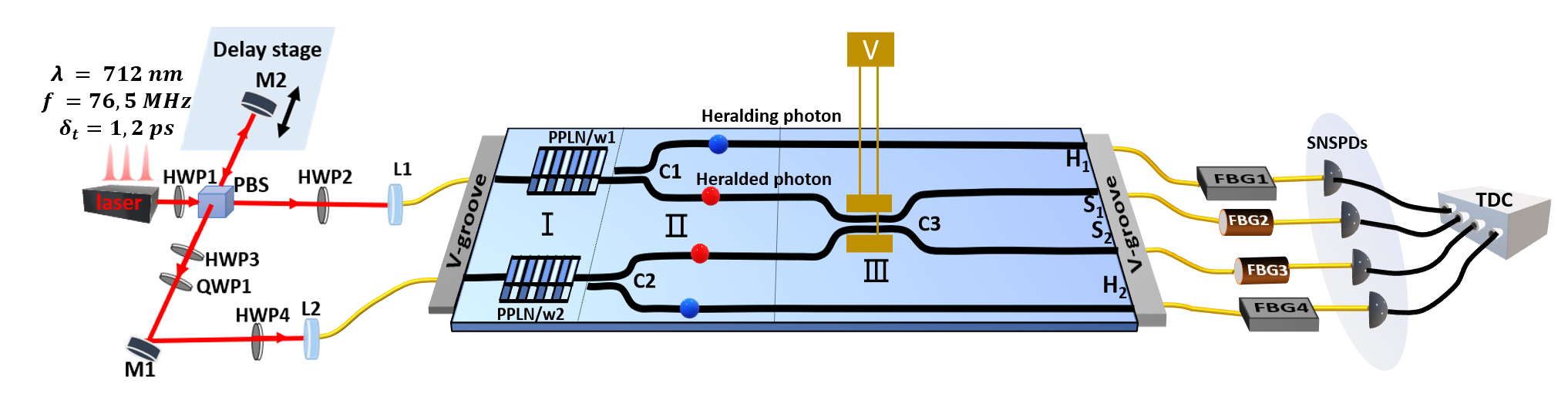}
\caption{Schematic of the monolithic LN chip with its associated pump, filtering and detection environment. The chip is partitioned into 3 regions, \uppercase\expandafter{\romannumeral1} for photon pair generation, \uppercase\expandafter{\romannumeral2} for photon pair splitting, and \uppercase\expandafter{\romannumeral3} for heralded entanglement manipulation. HWP: half-wave plate; QWP: quater-wave plate; PBS: polarizing beam splitter; M: mirror; L: lens; V-Groove: standard fiber array 127\,$\mu$m-spacing; C1/C2/C3: evanescently coupled waveguides; V: voltage; $H_1$/$H_2$: heralding modes; $S_1$/$S_2$: heralded modes; FBG: Fibre Bragg Grating filters; SNSPDs: superconducting nanowire single photon detectors; TDC: time-to-digital converter.\label{chip}}
\end{figure*}

Our device consists in a 4\,cm-long monolithic LN chip fabricated by soft-proton exchange~\cite{sebastien:02}. When operated appropriately, it aims at delivering heralded pairs of entangled photons at a telecom wavelength. As depicted in Figure~\ref{chip}, the operation principle relies on two integrated heralded single photon sources (HSPS) firing simultaneously, whose single photon outputs are routed toward a tunable coupler for entanglement manipulation by means of two-photon interference.
More precisely, the chip integrates several functions. In region~\uppercase\expandafter{\romannumeral1}, two periodically poled waveguides (PPLN/w) ensure the generation of photon pairs via efficient type-0 SPDC. Phase matching is engineered such that identical non-degenerate photon pairs are produced at the telecom wavelengths of 1310\,nm (signal) and 1560\,nm (idler) in each waveguide. In region~\uppercase\expandafter{\romannumeral2}, two evanescently coupled waveguides (C1 and C2) realize the wavelength demultiplexing of the photons for each pair toward four distinct spatial modes. In region \uppercase\expandafter{\romannumeral3}, signal photons are routed to the outer modes, $H_1$ and $H_2$, towards the heralding detectors. Their associated idler photons are routed to the inner modes, $S_1$ and $S_2$, towards the tunable coupler C3 whose splitting ratio can be tuned by means of electro-optical effect for quantum state engineering via controlled two-photon interference. When C3 is set to 50:50, destructive interference will lead to path entangled, so called N00N states, provided the interfering single photons are indistinguishable in terms of polarization, spatial mode and time of arrival. The pump laser is injected into the chip by means of a twin polarisation-maintaining-fibre array V-groove. Each input is independently adjustable in delay for the sole purpose of demonstration of two-photon interference. Fully integrated devices including a power splitter at the input of the chip would indeed exhibit non-adjustable matched optical paths showing no free parameter for proving high visibility HOM interference pattern. At the output of the chip, the photons are collected by a four single-mode-fibre array V-groove, and directed toward dedicated filtering and detection stages.

\subsection{Two sources of indistinguishable heralded pure single-photons}

In the context of quantum networks, where independent entanglement distribution segments are linked together via entanglement swapping, it has been shown that the picosecond regime offers an ideal trade-off in terms of near perfect two-photon interference and high photon-pair flux~\cite{Valivarthi:16,DAuria:20}. In addition, it allows exploiting ultra-dense wavelength division multiplexing systems with the UDWDM 25\,GHz telecommunication industry standard to obtain pure single photon states. Our device is therefore operated using a pump laser (Coherent Mira 900D), emitting picosecond pulses at 712\,nm, at a repetition rate of R=76.5\,MHz. The time synchronisation of pump pulses entering the chip is ensured, for demonstration purpose only, by a translating delay stage with femtosecond resolution.

\begin{figure}[htbp]
\centering
\includegraphics[width=1\linewidth]{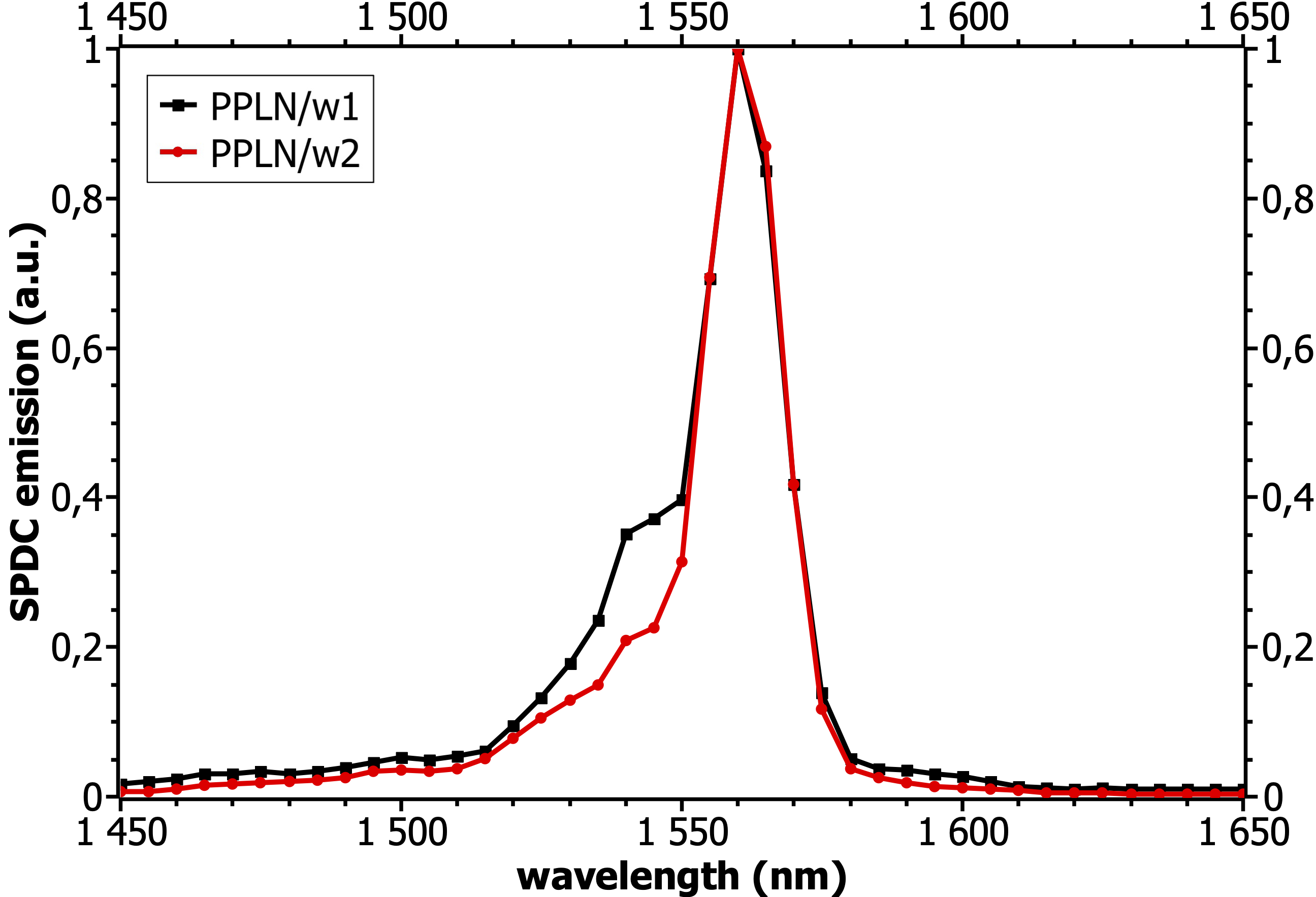}
\caption{Idler power spectra of the SPDC emission. \label{spectra}}
\end{figure}

For addressing photon pair generation in Region \uppercase\expandafter{\romannumeral1}, two single-mode PPLN waveguides with a poling period of $\Lambda$ = 13.1\,$\mu$m, have been engineered to produce non-degenerated vertically polarized paired photons at 1310\,nm and 1560\,nm from 712\,nm photons via type-0 SPDC process at the temperature of 373\,K. The source brightness has been measured to be of about $8\cdot10^8$ pairs.mW$^{-1}$.s$^{-1}$. Power spectra of the idler photons from both sources are shown in Figure~\ref{spectra} and exhibit near perfect overlap at 1560\,nm over a bandwidth of 20\,nm, in agreement with the theoretical expectation for a PPLN section of 1.6\,cm-long. By energy conservation, the signal photons are calculated to be at 1310\,nm over a similar bandwidth.

For our chip, the two-photon interference maximum visibility is directly related to the heralded single-photon purity $P=\sqrt{1-\left(\frac{\sigma_f^2}{\sigma_p^2+\sigma_f^2}\right)^2}$, where $\sigma_p$ and $\sigma_f$ are respectively the spectral bandwidth of the pump and interfering photons~\cite{Mosley:08}. The purity can also be measured through the time-integrated second-order correlation function $g^{(2)}(\tau)$ of the idler mode of our SPDC sources. In such a configuration the measured purity reads as $P=g^{(2)}(0)-1$~\cite{Eckstein:11}. In our case, the heralded single-photon purity is ensured by adequate filtering of the idler photons. Assuming a pump laser emitting 1.2\,ps pulses ($\sigma_p\sim 142\,$GHz), two 100\,GHz-DWDM FBGs at 1310\,nm and two 25\,GHz-UDWDM filters at 1560\,nm have been used for the signal and idler photons, respectively. Such bandwidths fit the conventional telecom standard (ITU) and allow realizing an excellent trade-off between heralded single photon purity and high photon-pair generation rate~\cite{Pierre:10,Meyer:17} since the theoretical purity of the idler mode is expected to be $\sim 99\%$. Second-order auto-correlation function at zero time delay $g^{(2)}(0)$ measurement has been performed on each of the non-heralded idler photons by using standard Hanbury-Brown \& Twiss setup simply exploiting the tunable coupler C3 set to 50:50 ratio. For the two PPLN/w sources, we measure $g^{(2)}(0) = 1.96 \pm 0.08$ and $1.93 \pm 0.07$, respectively, corresponding to source purities of $(96 \pm 8)\%$ and $(93 \pm 7)\%$.

\subsection{Wavelength demultiplexing isolation and overall losses}\label{loss}

The directional couplers C1 and C2 consist of two waveguides integrated close to each other over a given length~\cite{wooten:00}. If the spacing between the waveguides is sufficiently small, all the energy can be transferred from one to the other through evanescent coupling after a characteristic length which is wavelength dependent. Numerical simulations have been carried out using the beam propagation method (BPM) to optimize the spacing between the two waveguides and the associated coupling length in order to have the energy coupling of 1560\,nm and 1310\,nm in phase opposition.
Spectral separation of coupler C1 and C2 as well as the propagation losses have been measured using CW laser sources at 1560\,nm and 1310\,nm injected at the input of source 1 and 2. We report in Table~\ref{WDM} the performance of C1 and C2 for routing signal and idler photons as well as the optical transmission of the chip, i.e. the on-chip propagation losses.

\begin{table}[!htp]  
	\centering
	\caption{Performance of couplers C1 and C2 in terms of spectral separation and optical transmission of the chip. For instance, the losses associated with the propagation 1310\,nm photons from source 1 to the heralding mode ($H_1$) is 7\,dB while the isolation from being routed to signal mode ($S_1$)is 25.7\,dB.} 

		\begin{tabular}{|c|c|c|c|c|c|}
		\hline
		  \multirow{2}*{wavelength} & \multicolumn{2}{c|}{Source 1 (C1)} &  \multicolumn{2}{c|}{Source 2 (C2)} \\ 
		\cline{2-5}
		 & losses & isolation & losses & isolation \\
		\hline 
		 1310\,nm & 7\,dB & 25.7\,dB & 6.5\,dB  & 21.2\,dB  \\
		  1560\,nm & 7\,dB & 17.2\,dB & 6.2\,dB  & 14.7\,dB  \\
		\hline
	\end{tabular}\label{WDM}
\end{table}

The input-output optical transmission of the device, including on-chip propagation, photon collection in single mode fiber as well as the external filtering losses are measured to be about $\sim$ 14\,dB which mainly consists in on-chip propagation losses $\sim$ 7\,dB, chip-to-fiber coupling losses $\sim$ 3\,dB, and additional filtering stage losses $\sim$ 4\,dB.
Both couplers C1 and C2 show channel isolation of about 10\,dB at 1560\,nm and up to 15\,dB at 1310\,nm. Such a difference comes from the directional coupler design. A spacing separation of 11\,$\mu$m between the two waveguides ensures an approximate relation between the characteristic transfer lengths at 1310\,nm and 1560\,nm: $L_{c_{1310}} \approx 2L_{c_{1560}}$. In other words, when the photon pairs travel in the coupler of length $L$, the 1310\,nm photons are coupled to the other arm, while the photons at 1560\,nm are coupled back and forth to the original arm. Consequently, 1560\,nm photons remain mostly in the input waveguide, whereas 1310\,nm photons cross to the coupled waveguide. Hence, such a design guarantees a good isolation for both wavelengths, but not an optimal one for both. A better isolation on 1310\,nm photons is therefore a matter of fabrication tolerance.

\subsection{Two-photon state manipulation}
\begin{figure}[!h]
\centering
\includegraphics[width=1\linewidth]{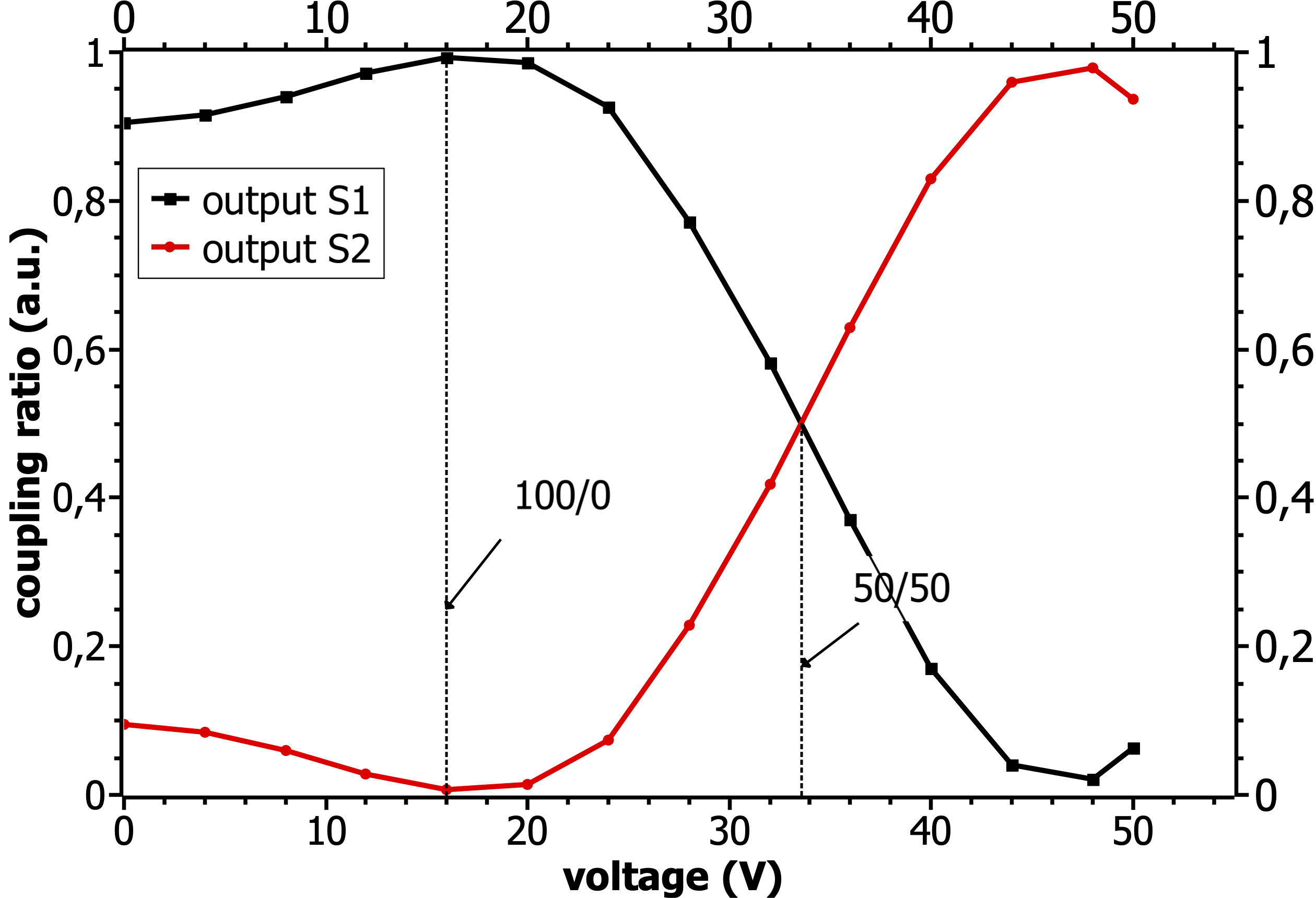}
\caption{Coupling ratio calibration of C3 as a function of the applied votage. Laser light at 1560\,nm is injected through source PPLN/w1 and measured at output $S_1$ and $S_2$. The dashed lines figure out two extreme cases: $16$\,V for 100:0, $34$\,V for 50:50 operation.\label{EO_coupler}}
\end{figure}

Coupler C3 enables controlling the two-photon interference via electrical tuning of the coupling ratio. Technically speaking, it is similar to C1 and C2 described earlier in a configuration where all the energy is fully transferred from one waveguide to the other. Afterwards, controllable decrease of the coupling ratio is achieved by detuning the propagation constant between the two waveguides by means of the nonlinear electro-optical effect. Lumped electrodes, deposited on top of the two coupled waveguides with a silica buffer layer, are used to tune the coupling ratio from 0:100 (transmission) to 100:0 (reflection). To remove the short term DC drift due to current leakage through the buffer layer~\cite{Yamada:81}, the voltage applied between the electrode pads is switched continuously between +V and -V at a frequency of 1\,kHz by using a squared shape. Meanwhile, we verify that the transient voltage has low impacts on the chip performances. To this end, we assess the performances of the couplers by using CW laser at 1560\,nm injected in one input at a time and then measure output mode powers $S_1$ and $S_2$ as a function of the applied voltage. A coupling ratio smoothly ranging from 100:0 to 0:100 is shown in Figure~\ref{EO_coupler}. The specific point of 50:50 operation is determined for an applied voltage of 34\,V. Such a voltage is easily accessible using a standard function generator coupled to low-level amplifier.

\section{Quantum regime}
\hspace*{1em} Figure~\ref{chip} shows the schematic of the setup utilized for the 4-fold coincidence measurement related to the heralded generation of path entangled states. Here, we analyze the two-photon states in mode $S_1$ and $S_2$, which have interfered on chip, whenever two heralding signal photons have been detected out of modes $H_1$ and $H_2$. Each of the four optical outputs is connected to superconducting detectors (ID 281) showing $\sim$80\% detection efficiency and a dark count rate of about 200\,Hz. Each detector is connected to a time-tagging module (HydraHarp 400) whose resolution is of 10\,ps. 

We start by setting the voltage driving C3 to 34\,V. Two-photon interference visibilities are measured and obtained by changing the optical delay between photons $S_1$ and $S_2$. This physical situation corresponds to the well-known Hong-Ou-Mandel interference and allows the purity of the prepared N00N state~\cite{Menssen:17} to be estimated. The 4-channel time-tagged data sets are analyzed as a function of the input path-length difference by moving the mirror M2 with a motorized translation stage. In a second time, we also report the coincidence rate as a function of the voltage (two extreme cases V = 34\,V and 18\,V are emphasized) applied to the tunable coupler when the input path-lengths are matched. Such a configuration corresponds to the aimed operating mode of the chip when the user only controls the coupler voltage to tune the quantum state of the heralded pairs. 

\subsection{Hong-Ou-Mandel dip measurement}

\begin{figure}[!h]
\centering
\includegraphics[scale=0.5]{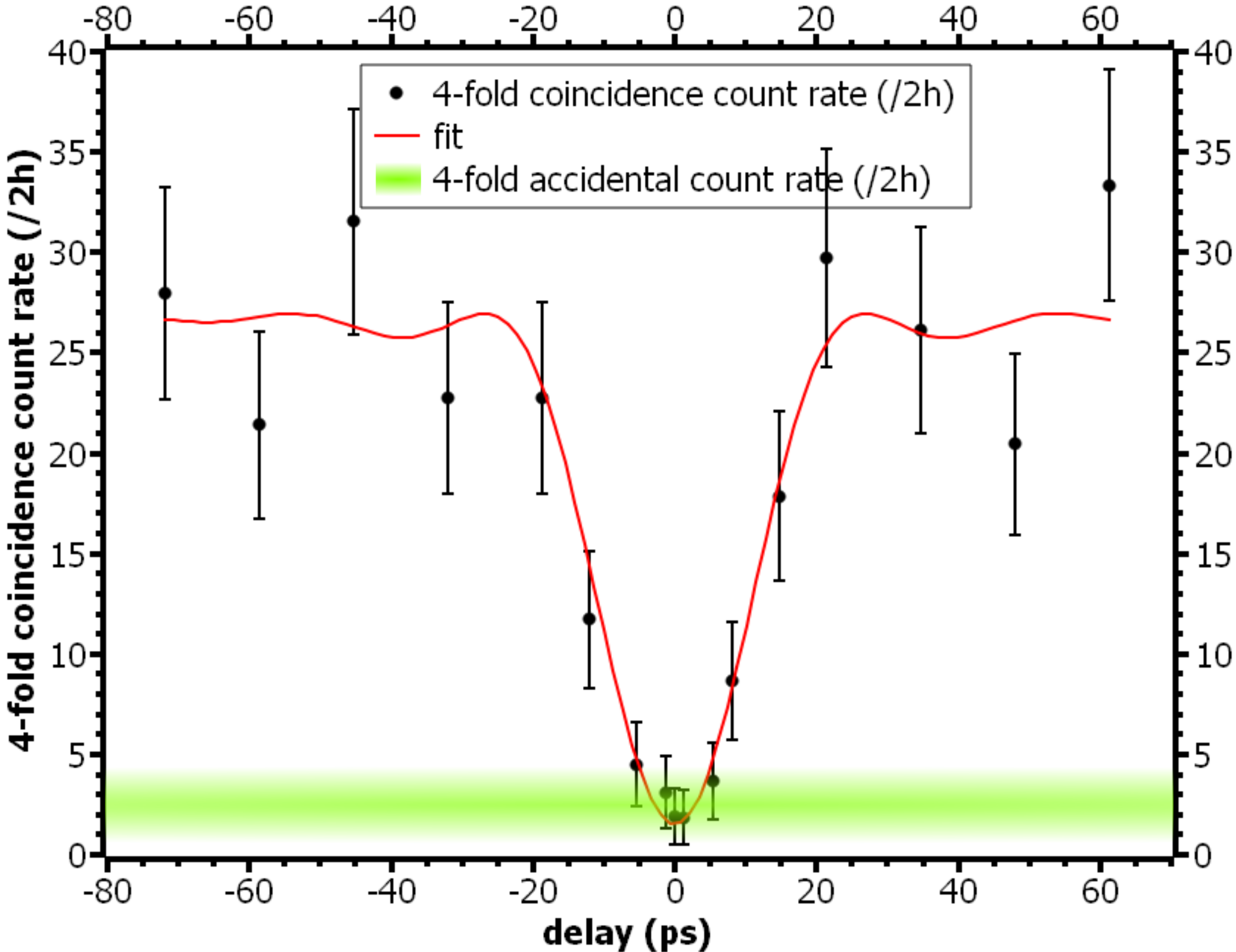}
\caption{Raw 4-fold coincidence count rate as a function of the relative delay in picoseconds between photons $S_1$ and $S_2$ for an integration time of 2 hours. The multipair contribution is the principal noise factor and is reported on the graph in green area. The error bars are assumed as $\sqrt N$ due to the poissonian statistic of photon pair detections ($N$ represents the measured coincidence count).\label{HOM}}
\end{figure}
 
We inject the chip with $\sim$ 0.7\,mW of pump power, which translates into a mean number of $9\cdot10^{-3}$ pairs of photon per pulse, ensuring low multi-photon contribution but still allowing a reasonable four-fold coincidence rate, when taking into account the overall losses~\cite{Smirr:11}. We set the coupler C3 to 50:50 (i.e. 34\,V) and then integrate four-fold coincidences over 2 hours for different path-delays. The results are presented in Figure~\ref{HOM} and have been fitted using a sinc$^2$ function accordingly to the "squared" spectral wavepacket of our heralded photons in channels $S_1$ and $S_2$. We use only two free parameters corresponding to the visibility $V$ and mean amplitude. The width of the HOM dip is estimated to be of about 7.6\,mm, compatible with the bandwidth of the interfering photons.

According to the fitting function, we obtain a raw visibility of $94\%$. The dark count contribution to four-fold accidentals is negligible but the optical noise due to multipair contributions has been measured to contribute up to 3 coincidences per 2h corresponding to the shaded area in Figure~\ref{HOM}. To estimate the induced 4-fold accidental coincidence count, we record the energy-mismatched pair contributions to the 4-fold rate by changing the central wavelength of filters $H_1$ and $H_2$ from 1310\,nm to 1314\,nm. Most of the non-perfect visibility is accounted to those multipair contributions.

\subsection{Demonstration of tunable source of heralded entanglement}

\begin{figure*}[!htb]
\centering
\begin{tabular}{c c}
  \includegraphics[width=0.45\linewidth]{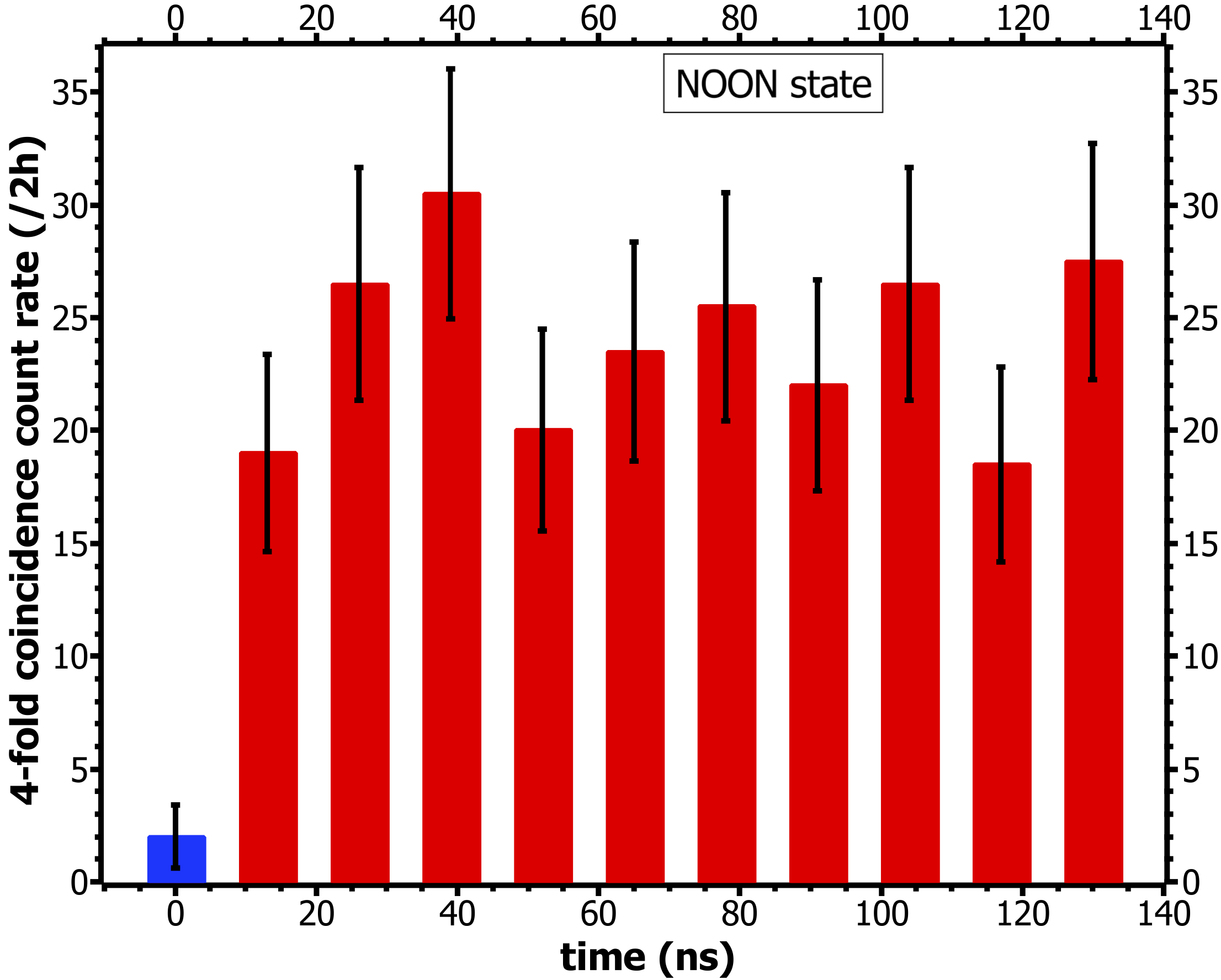} & \includegraphics[width=0.45\linewidth]{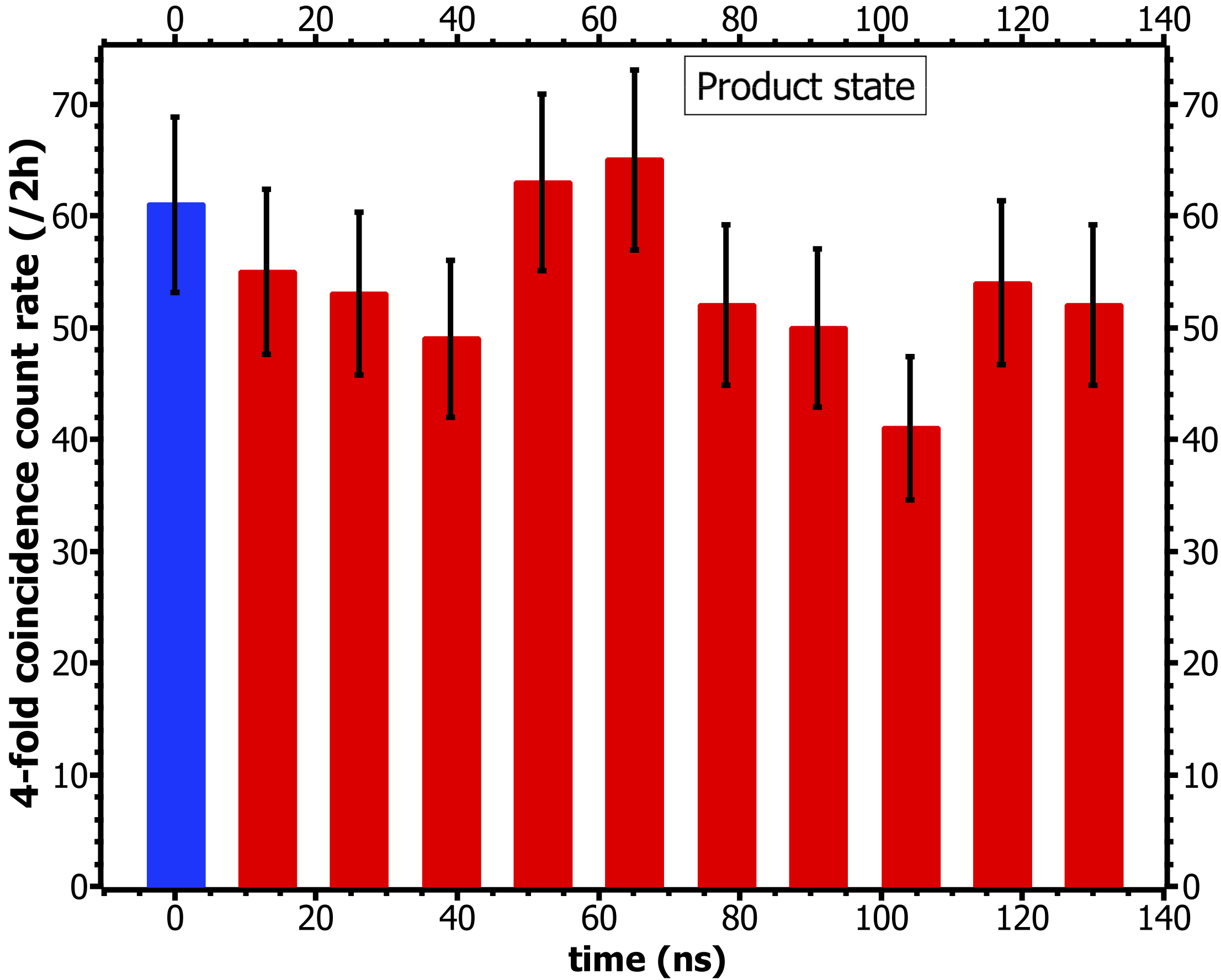}\\
  (a) & (b)
\end{tabular}
\caption{4-fold coincidence count rate measurement for relative delay = 0\,ps as a function of the pulse offset in time. (a) V = 34\,V, N00N state case, and (b) V = 18\,V, product state case. The first peak shows 4-fold coincidence counts for two pairs of photon simultaneously created, and the following peaks are for two pairs of photon created with a time interval $n\times\delta t$ (n = 1, 2, 3, ...), $\delta t$ being the repetition time of pump laser which is equal to 13\,ns. The error bars are assumed to be $\sqrt N$ according to the poissonnian statistics of the 4-fold coincidences. The reduced number of 4-fold coincidences for $\Delta T\neq 0$ in the N00N case can be explained by the extra 3\,dB loss of the 50:50 coupler.\label{product}}
\end{figure*}

In Figure~\ref{product}, we show the experimental 4-fold coincidence count rate histograms as a function of the pulse offset in time for 2 extreme cases: N00N and product states, respectively. The blue colored bars at $t=0$ correspond to heralded pairs of simultaneous photons whose state can be engineered from product state to path-entangled N00N state. All the subsequent red bars correspond to heralded pairs whose photons are delayed by $N$ times the laser pulse interval and can be seen as coherent superpositions of "early" and "late" emission times. 
Based on the overall loss budget, when the source generates product states (i.e. the coupler C3 is set to 100:0), we estimate the 4-fold coincidences rate $R_{1234}$ to reach about $\sim59$ events per 2 hours. It reads:
\begin{equation}
\centering
R_{1234} = R*\bar n_p^2*\mu_1*\mu_2*\mu_3*\mu_4*\eta^4*T,\label{rate}
\end{equation}
where $R$ is the pump pulse repetition rate of 76.5\,MHz, $\bar n_p$ is the mean number of pairs per pulse, $\mu_i$ are the associated losses on each channel 1,2,3 and 4 calculated at section~\ref{loss}, while $\eta$ is the detection efficiency of the SNSPDs and $T$ represents the integration time.

In this configuration, an interesting metric corresponds to the heralding rate which translates into the number of times the source has emitted a two-photon state usable for any other application such as communication or computation. Experimentally, we obtained an heralding rate of $\sim$200\,Hz leading to $\sim$60 detections per 2h of separated states, well in accordance with the numerical estimation of eq.\ref{rate} . In Figure~\ref{product}a, the reported 4-fold rate is twice lower than expected for when $\Delta T\neq 0$ and is fully accounted to the extra 3\,dB loss of the coupler in the 50:50 mode.

\section{Conclusion}
\hspace*{1em}We have demonstrated heralded two-photon states generation and manipulation on monolithic lithium niobate chip. Such a device aims at generating reconfigurable heralded path-coded two-photon states. It gathers five different integrated elements to enable successively photon pair production, wavelength demultiplexing of the pairs and two-photon interference on a tunable coupler. Each optical function of the chip shows high-level performances in terms of losses and efficiency. The monolithic configuration ensures high stability, low power consumption, and low operating voltage. 
Exploiting telecom compatible components, the quality of the produced entangled state has been shown by means of Hong-Ou-Mandel interference exhibiting a raw visibility of $94\%$ only limited by the multipair contributions. In nominal operating conditions, the source exhibits an heralding rate of entangled photons of $\sim 200$\,Hz with quantum state switching from product $|1,1\rangle$ to N00N $\frac{1}{\sqrt{2}}\left(|2,0\rangle+|0,2\rangle\right)$ states in less than 1\,ms by applying about $\sim 34$\,V. The high brightness as well as entanglement quality, and the alignment-free operation make our device a valuable candidate for quantum applications. Remarkably to note that, exploiting a telecom optical clock as in~\cite{DAuria:20}, the rate could be extended by one order of magnitude while maintaining the quantum state quality.

Such a chip finds natural application in quantum networking whose heralding signal provide meaningful information for repeater-based links. Beyond quantum communication, the architecture proposed for this chip can be conveniently used in the context of continuous variables (CV) quantum optics. Depending on the input pumping power at 712\,nm, the SPDC stages generate photon pairs, as for this work, or bright squeezed states. In CV regime, at higher SPDC pump powers, a similar scheme has been used to generate two single-mode or one two-mode squeezed state in continuous wave regime~\cite{Lenzini:18}. Possible extension would also cover the conditional preparation of odd/even Schrödinger kittens in the mode $H_1$ and $H_2$ of the chip, via heralded single photon substraction toward coupler C3 and further produce a coherent superposition of heralded photon subtraction. Based on the choice of their relative phase, it is possible to use this detection signal to herald a strongly entangled non-local superposition of coherent states~\cite{Ourjoumtsev:09}. This latter represents a resource useful for long distance quantum teleportation and phase-space rotations of coherent qubits. Eventually, such an architecture can be equally used to prepare in a heralded fashion an hybrid DV-CV entangled state with single-rail encoding on its DV part~\cite{Morin:14}. This can be achieved by mixing on C3 a photon from a pair generated by the upper SPDC and a single photon subtracted from a squeezed state from the lower SPDC source.

\section{Disclosures}

\noindent The authors declare no conflicts of interest.

\section{Acknowledgment}
In loving memory of Marc Pierre De Micheli. We thanks Leonid Krivitskiy and Bhaskar Kanseri for fruitful discussions and help.
X.Hua acknowledges PhD funding from R\'egion Provence Alpes C\^ote d'Azur and University C\^ote d'Azur.
This work has been conducted within the framework of the OPTIMAL project, funded by the European Union by means of the Fond Europ\'een de d\'eveloppement regional (FEDER). The authors acknowledge financial support from Agence Nationale de la Recherche (ANR) through the INQCA (ANR-14-CE26-0038) project, Conseil R\'egional PACA through the SNATCH project (Apex 2019), the Universit\'e C\^ote d’Azur (PEANUTS and CSI 2018) and the French government through its Investments for the Future programme under the Universit\'e C\^ote d’Azur UCA-JEDI project (Quantum@UCA) managed by the ANR (ANR-15-IDEX-01).


\end{document}